\documentclass[sigconf]{acmart}
\AtBeginDocument{}

\copyrightyear{2025}
\acmYear{2025}
\setcopyright{cc}
\setcctype{by}
\acmConference[MM '25]{Proceedings of the 33rd ACM International Conference on Multimedia}{October 27--31, 2025}{Dublin, Ireland}
\acmBooktitle{Proceedings of the 33rd ACM International Conference on Multimedia (MM '25), October 27--31, 2025, Dublin, Ireland}\acmDOI{10.1145/3746027.3755190}
\acmISBN{979-8-4007-2035-2/2025/10}
\acmDOI{10.1145/3746027.3755190}

\usepackage{pifont}
\usepackage{multirow}
\usepackage{hyperref}

\begin{document}

\title{PESTalk: Speech-Driven 3D Facial Animation with Personalized Emotional Styles}

\author{Tianshun Han}
\orcid{0009-0004-3393-1597}
\email{3230002542@student.must.edu.mo}

\author{Benjia Zhou}
\orcid{0000-0003-4883-5552}
\email{zhou.benjia0228@gmail.com}

\affiliation{
  \institution{Macau University of Science and Technology}
  \city{Macau}
  \country{China}}

\author{Ajian Liu}
\orcid{0000-0002-7788-9368}
\affiliation{
  \institution{MAIS, CASIA}
  \city{Beijing}
  \country{China}}
\affiliation{
  \institution{Macau University of Science and Technology}
  \city{Macau}
  \country{China}}
\email{ajian.liu@ia.ac.cn}

\author{Yanyan Liang}
\orcid{0000-0002-5780-8540}
\affiliation{
  \institution{Macau University of Science and Technology}
  \city{Macau}
  \country{China}}
\email{yyliang@must.edu.mo}

\author{Du Zhang}
\orcid{0000-0002-8301-2706}
\affiliation{
  \institution{Macau University of Science and Technology}
  \city{Macau}
  \country{China}}
\email{duzhang@must.edu.mo}

\author{Zhen Lei}
\orcid{0000-0002-0791-189X}
\affiliation{
  \institution{MAIS, CASIA}
  \city{Beijing}
  \country{China}}
\affiliation{
  \institution{Macau University of Science and Technology}
  \city{Macau}
  \country{China}}
\email{zhen.lei@ia.ac.cn}

\author{$\text{Jun Wan}{}^{\dagger}$}\thanks{$\dagger$ Corresponding Author.}
\orcid{0000-0002-4735-2885}
\affiliation{
  \institution{MAIS, CASIA}
  \city{Beijing}
  \country{China}}
\affiliation{
  \institution{Macau University of Science and Technology}
  \city{Macau}
  \country{China}}
\email{jun.wan@ia.ac.cn}

\renewcommand{\shortauthors}{Tianshun Han et al.}

\begin{abstract}
Speech-driven 3D facial animation aims to synthesize realistic emotional facial expressions that match the input speech. However, existing approaches are constrained by two key limitations: (1) These methods rely on pre-trained models (\textit{e.g.}, Wav2Vec 2.0) as audio emotion feature extractors, which neglect critical frequency-domain characteristics, thereby emphasizing the challenge of discriminating between similar emotion categories. (2) They treat audio emotions as generic categorical states, ignoring individual differences in emotional expression, ultimately producing over-smoothed emotional representations that appear repetitive and stereotypical. To that end, we introduce \textbf{PESTalk}, a novel approach that generates 3D facial animations with \textbf{P}ersonalized \textbf{E}motional \textbf{S}tyles directly from speech inputs, thus significantly enhancing the realism of facial animations. Specifically, since acoustic frequency cues contain essential emotional information, we first propose a \textbf{D}ual-\textbf{S}tream \textbf{E}motion \textbf{E}xtractor (\textbf{DSEE}), which captures both time-domain variations and frequency-domain characteristics of audio signals to extract fine-grained affective features and subtle emotional nuances. Furthermore, we design an \textbf{E}motional \textbf{S}tyle \textbf{M}odeling \textbf{M}odule (\textbf{ESMM}) to achieve personalized emotional styles. This module first establishes a baseline representation for each subject based on voiceprint characteristics, then progressively refines it by continuously integrating emotional features. Ultimately, this process constructs a personalized emotional style representation for each subject in each emotion category, capturing their unique expression patterns. Finally, considering the scarcity of the 3D emotional talking face data, we employ an advanced facial capture model to extract pseudo facial blendshape coefficients from 2D emotional data, thereby constructing a large-scale 3D emotional talking face dataset with diverse emotions and personalized expressions (3D-EmoStyle). Extensive quantitative and qualitative evaluations show that PESTalk can generate realistic 3D facial animation and outperform state-of-the-art methods. The codes and dataset are available at: \textit{\url{https://github.com/tianshunhan/PESTalk}}.

\end{abstract}



\maketitle
\section{Introduction}
Speech-driven 3D facial animation aims to generate realistic facial movements for 3D characters based on given speech input. Although there have been substantial advancements~\cite{fan2022faceformer,peng2023selftalk,park2023said,stan2023facediffuser, han2024pmmtalk,lin2024glditalker} in lip synchronization, they neglect the overall facial expressions, resulting in only reliable lip movement. Consequently, the current researches~\cite{emotalk,shen2024deitalk} is shifting towards creating comprehensive facial animations. Such animations are crucial for enhancing user experience, particularly in interactions with non-player characters in video games or emotionally responsive virtual chatbots~\cite{kim2024deeptalk}.

\begin{figure*}[h]
\centering
  \centering 
  \includegraphics[width=0.95\textwidth]{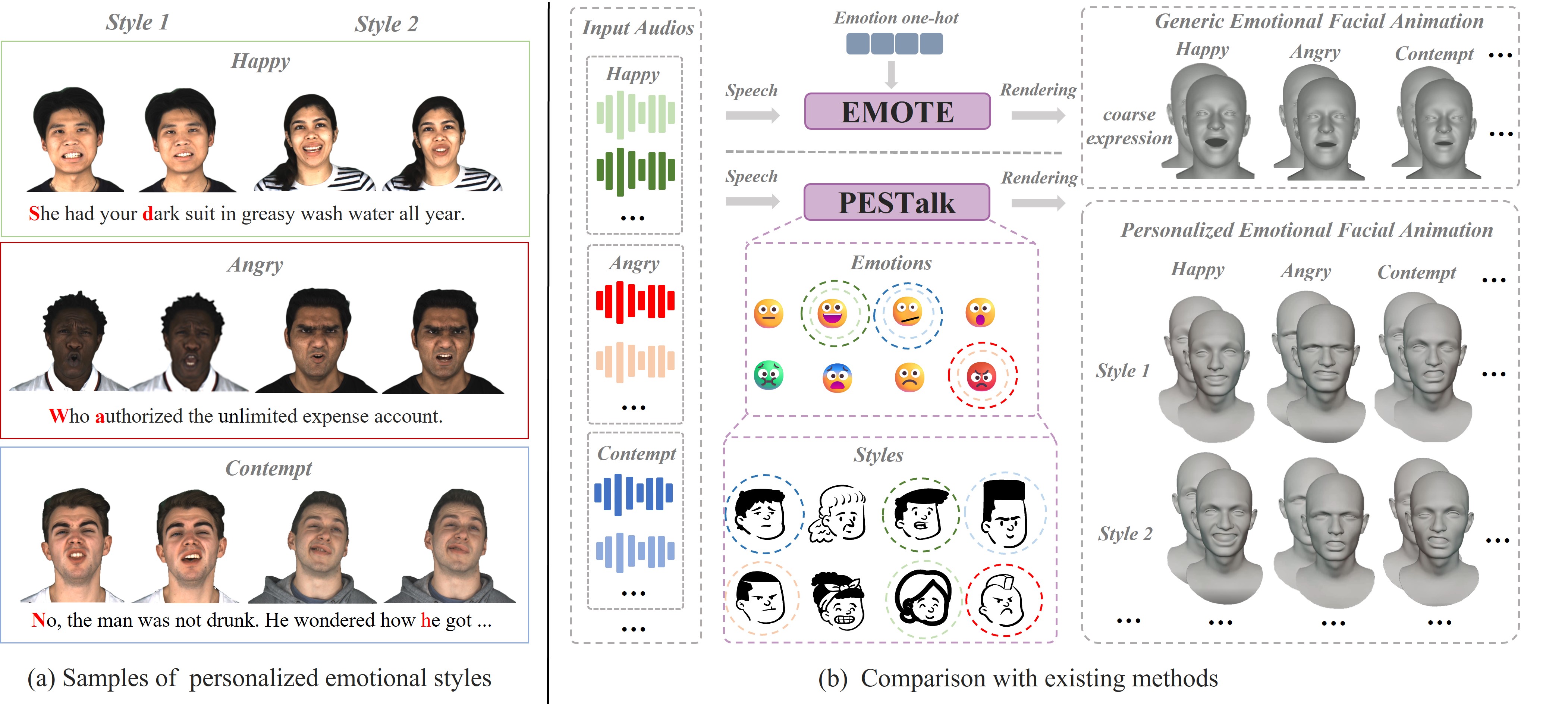}
  \caption{(a) When expressing the same emotion with identical sentences, people show different facial expression patterns. This is influenced by habitual behaviors and cultural backgrounds, known as personalized emotional styles. (b) Prior work (\textit{e.g.}, \textsc{EMOTE}~\cite{danvevcek2023emotional}) treats emotion as one-hot embeddings, producing averaged expressions, while PESTalk dynamically selects optimal expressions by analyzing emotions and voiceprints, generating realistic 3D animations with personalized styles.}
  \label{fig1}
\end{figure*}

However, recent methods~\cite{emotalk,danvevcek2023emotional,shen2024deitalk,haque2023facexhubert,karras} still suffer from two major limitations. 
First, most speech-driven 3D studies~\cite{emotalk,danvevcek2023emotional,shen2024deitalk} rely on pre-trained speech models like Wav2Vec 2.0~\cite{baevski2020wav2vec} to extract emotional information from input speech. Nevertheless, this approach relies on temporal convolutions to extract time-domain features, obscuring cross-frequency band correlations, thereby limiting its ability to discriminate between similar emotional categories~\cite{li2022deep,matsumoto2013culture}. Moreover, some works~\cite{danvevcek2023emotional,wang20213d,karras} treat audio emotions as generic categorical states and represent them using one-hot vectors. Although this strategy offers an efficient way to regulate emotional expression in animation, it overlooks the personalized emotional expression, leading to over-smoothed facial animations. In reality, as illustrated in Fig.~\ref{fig1} (a), different speakers exhibit unique facial dynamics even when conveying the same emotion, highlighting that emotional expression is highly individualized~\cite{li2022deep,matsumoto2013culture}. 

To address this challenge, we propose \textbf{PESTalk} (shown in Fig.~\ref{fig1} (b)), a novel speech-driven 3D facial animation model that extracts emotional information and personalized style cues from audio signals to generate highly realistic 3D facial expressions. Our framework comprises two essential modules: the Dual-Stream Emotion Extractor (DSEE), designed to enhance the capability of model in extracting fine-grained emotional features, and the Emotional Style Modeling Module (ESMM), which enables personalized emotional styles generation. Specifically, the DSEE consists of two independent audio encoders: a temporal stream encoder based on Temporal Convolutional Networks (TCN) to capture time-domain variations, and a frequency stream encoder utilizing mel-spectrogram analysis to extract frequency-domain characteristics. This complementary dual-stream mechanism enables a more comprehensive capture of subtle acoustic cues in emotional expression. Additionally, we develop the ESMM to enhance personalized expressiveness. The module first establishes baseline expression templates using speaker voiceprint characteristics, then dynamically constructs a two-dimensional \textit{\textless speaker $\times$ emotion\textgreater} emotional style library through continuously integrating emotional features of multiple emotional samples. During inference, the ESMM automatically retrieves the most matching personalized emotional style features from the input audio, thereby enabling the model with personalized expression capabilities. Finally, based on the biomechanical distinction that mouth movements (high-frequency for speech articulation) and upper-face motions (low-frequency for emotional expressions) require different processing, a partitioned style-guided fusion decoder is proposed to effectively handle this disparity. 

To train the proposed network, emotional speeches with various personalized styles paired with corresponding 3D facial expressions are required. However, the absence of publicly suitable datasets poses a new challenge. To tackle this issue, we propose an emotional 3D talking face dataset, termed the 3D-EmoStyle dataset. To build this dataset, we use an advanced facial capture
model~\cite{lugaresi2019mediapipe} to extract "pseudo" 3D facial blendshape labels from 2D audio-visual datasets~\cite{zhang2021flow,wang2020mead}. Furthermore, to adapt the mesh-based algorithms, we employ deformation transfer~\cite{sumner2004deformation} to create 52 FLAME head templates from the facial blendshape coefficients, allowing for efficient conversion between various facial animations.

In summary, our primary contributions are as follows:
\begin{itemize}
\item We propose a novel approach (PESTalk) to generate highly realistic 3D facial animation with personalized emotional styles, presenting state-of-the-art performance.
\item We introduce a dual-stream emotion extractor (DSEE) to extract fine-grained emotional features and subtle nuances.
\item We design an emotional style modeling module (ESMM) to enable personalized emotional styles generation.
\item We collect a 3D talking face dataset (3D-EmoStyle) with diverse emotion categories and personalized expressions, including both facial blendshape and mesh vertex coefficients.
\end{itemize} 

\section{Related Work}
\subsection{Speech-Driven 3D Facial Animation}

Previously, 2D facial animation methods~\cite{sheng2023towards,ma2024cvthead,nazarieh2024portraittalk,zhong2023identity,zheng2024memo} have garnered significant attention, primarily focusing on image-driven or speech-driven approaches to generate realistic videos of speaking individuals. However, these methods are not directly applicable to 3D character models, which are widely used in 3D games. As a result, speech-driven 3D facial animation~\cite{park2023said,emotalk,shen2024deitalk,stan2023facediffuser,chen2023diffusiontalker} has recently emerged as a prominent research direction. 

VOCA~\cite{voca} pioneers the use of CNNs to map audio signals to 3D meshes, enabling the generation of facial animations with diverse speaking styles. MeshTalk~\cite{richard2021meshtalk} advances the field by focusing on full-face animations, successfully disentangling speech-correlated and speech-uncorrelated facial motions. FaceFormer~\cite{fan2022faceformer} introduces a transformer-based model to generate continuous facial movements in an autoregressive manner, setting a new benchmark for temporal coherence. CodeTalker~\cite{xing2023codetalker} further enriches the field by modeling the facial motion space using discrete primitives, achieving highly expressive animations. More recently, FaceDiffuser~\cite{stan2023facediffuser} becomes the first to apply probabilistic diffusion models to generate multiple facial motion options for a given speech, emphasizing lip-sync accuracy. 

Despite these advancements, a critical limitation still persists. Current methods mainly concentrate on lip synchronization while overlooking the natural expression of emotions through facial animations. Capturing the subtle emotional nuances conveyed in speech is of great significance for enhancing the realism of 3D facial animations. In contrast, our work does not merely focus on lip synchronization but is also capable of achieving emotional expressions across the entire face.  

\subsection{Emotional 3D Facial Animation}
The high coupling of content, emotion, and style in audio makes their decoupling a key challenge in realistic facial animation generation, drawing much research interest~\cite{emotalk,fu2024mimic,shen2024deitalk,danvevcek2023emotional}. EmoTalk~\cite{emotalk} introduces an emotion disentanglement mechanism to extract emotional information from speech input, but its reliance on curated training data limits its effectiveness in zero-shot settings. DEITalk~\cite{shen2024deitalk} addresses emotional saliency variations in long audio contexts by designing a dynamic emotional intensity module and a dynamic positional encoding strategy. EMOTE~\cite{danvevcek2023emotional} employs a content-emotion exchange mechanism to supervise diverse emotions on the same audio while maintaining precise lip synchronization. Mimic~\cite{fu2024mimic} advances the field by learning disentangled representations of speaking style and content through two distinct latent spaces, enabling detailed modeling of 3D facial expressions.

Even with these developments, current methods often focus on facial emotional expressions while neglecting personalized emotional styles, resulting in less realistic animations. Additionally, most approaches are mesh-based, directly mapping speech signals to face mesh vertex coordinates, which can be challenging for artists and difficult to integrate into existing workflows. In this work, we introduce a novel speech-driven 3D facial animation model that extracts emotional and style cues from audio to generate highly realistic expressions. By using facial blendshapes as the model output, we ensure seamless integration into industrial pipelines and enhance usability for professional animators.

\begin{figure*}[h]
\centering
\includegraphics[width=0.95\textwidth]{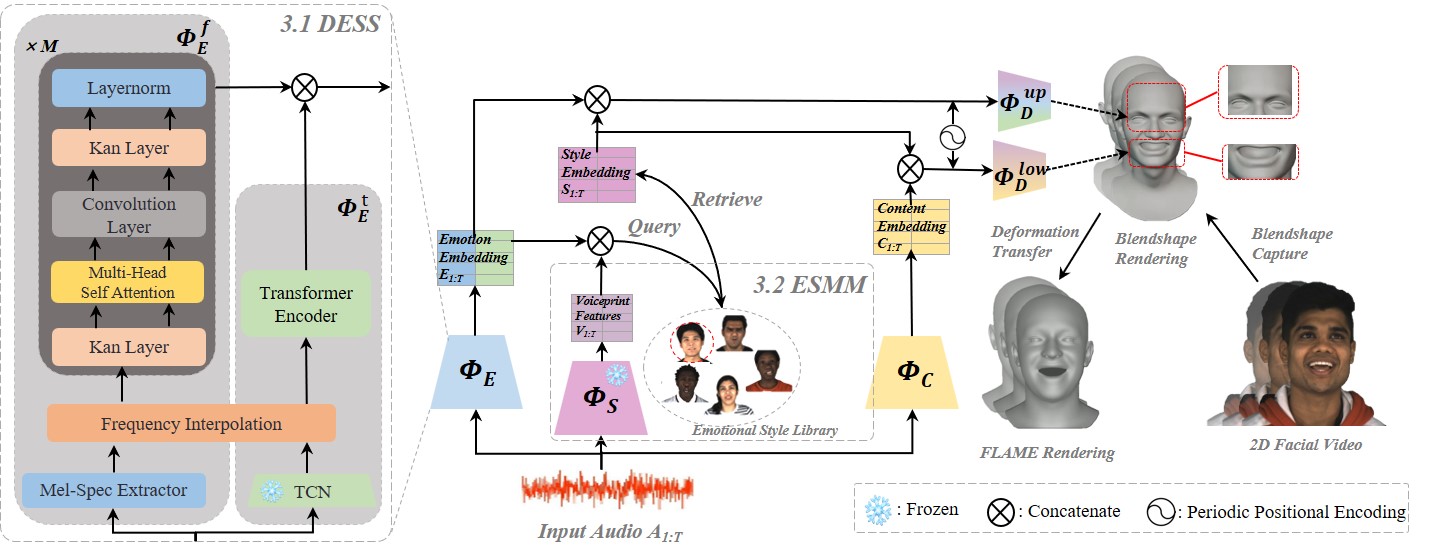}
\caption{Overview of PESTalk. Given a speech input ${A}_{1: T}$, PESTalk extracts emotion features $E_{1: T}$ and voiceprint information $V_{1: T}$. Then, it uses these two types of information to query the emotional style library and match the closest personalized emotional style $S_{1: T}$. After that, PESTalk extracts content features $C_{1: T}$ from the speech and integrates these features. Subsequently, two distinct sets of decoders, $\boldsymbol{\Phi}_{D}^{up}$ and $\boldsymbol{\Phi}_{D}^{low}$, combine these integrated features to generate facial blendshape coefficients for the upper and lower face, respectively. Additionally, these coefficients also can be used to animate a FLAME model.}
\Description{Overview of PESTalk. Given a speech input ${A}_{1: T}$, PESTalk extracts emotion features $E_{1: T}$ and voiceprint information. Then, it uses these two types of information to query the emotional style library and match the closest personalized emotional style $S_{1: T}$. After that, PESTalk extracts content features $C_{1: T}$ from the speech and integrates these features. Subsequently, two distinct sets of decoders, $\boldsymbol{\Phi}_{D}^{up}$ and $\boldsymbol{\Phi}_{D}^{low}$, combine these integrated features to generate facial blendshape coefficients for the upper and lower face, respectively. Additionally, these coefficients also can be used to animate a FLAME model.}
\label{method}
\end{figure*}

\section{Method}
\textbf{Overview.} In this section, we detail the architectural components of our proposed framework, PESTalk (shown in Fig.~\ref{method}), which generates highly realistic 3D facial animations with personalized emotional styles directly from speech inputs. The framework consists of two key modules: the Dual-Stream Emotion Extractor (DSEE, discussed in Section 3.1), which enhances the ability to extract fine-grained emotional features, and the Emotional Style Modeling Module (ESMM, covered in Section 3.2), which facilitates the generation of personalized emotional styles. Next, a style-guided fusion decoder (detailed in Section 3.3) transforms these features into facial animations. Additionally, we introduce a novel pairwise disentanglement mechanism (detailed in Section 3.4) to effectively separate content and emotion features of the input speech. The following subsections elaborate on the detailed design of each module.

\textbf{Formulation.} Let $\boldsymbol{B}_{1: T}=\left(\boldsymbol{b}_{1}, \ldots, \boldsymbol{b}_{T}\right)$, $\boldsymbol{b}_{t} \in \mathbb{R}^{52}$ be a $T$-length sequence of facial blendshape coefficients, which describes the ground truth of 3D face movements. Let ${A}_{1: T}=\left({a}_{1}, \ldots, {a}_{T}\right)$ be a sequence of speech snippets, and each ${a}_{t} \in \mathbb{R}^{D}$ has $\boldsymbol{D}$ samples to align with the corresponding (visual) frame $\boldsymbol{b}_{t}$. The proposed model takes speech ${A}_{1: T}$ as input, predicting facial blendshape coefficients $\hat{\boldsymbol{B}}_{1: T}=\left(\hat{\boldsymbol{b}}_{1}, \ldots, \hat{\boldsymbol{b}}_{T}\right)$. Formally,
\begin{equation}
\hat{\boldsymbol{b}}_{t}=\operatorname{PESTalk}_{\theta}\left({a}_{t}\right),
\end{equation}
where $\theta$ indicates the model parameters, $t$ is the current time-step in the sequence and $\hat{\mathbf{b}}_{\mathbf{t}} \in \hat{\mathbf{B}}_{\mathbf{1:T}}$. 

\subsection{Dual-Stream Emotion Extractor}
Prior works~\cite{emotalk, danvevcek2023emotional, kim2024deeptalk} primarily utilize pre-trained speech models as emotion encoders, which fail to adequately capture frequency-domain characteristics and often confuse similar emotional categories~\cite{li2022deep,matsumoto2013culture}. To overcome this limitation, we propose a novel dual-stream emotion extractor $\boldsymbol{\Phi}_{E}$ (see Fig. 2 left) that synergistically models both time-domain and frequency-domain emotional cues. The temporal stream $\boldsymbol{\Phi}_{E}^{t}$ is a pre-trained audio model fine-tuned on emotion~\cite{baevski2020wav2vec}. It is composed of several frozen TCN layers and multi-layer transformer blocks to extract robust time-domain features. The frequency stream $\boldsymbol{\Phi}_{E}^{f}$ incorporates a KAN-enhanced Conformer architecture that replaces traditional MLPs with KAN layers~\cite{liu2024kan} which employs spline-parameterized activation functions, enabling more precise modeling of complex emotional patterns in the frequency domain. This hybrid architecture simultaneously captures both long-term dependencies and short-term variations.

Specifically, for an input audio segment ${A}_{1: T}=\left({a}_{1}, \ldots, {a}_{T}\right)$, we first extract time-domain features through TCN layers and compute mel-spectrogram features. After interpolation transformations, we process them through parallel pathways: the frequency features undergo KAN-Conformer blocks to extract $E_{f_{1:T}}$, while the temporal features pass through a transformer encoder to obtain $E_{t_{1:T}}$. These are concatenated $[E_{t_{1:T}} \parallel E_{f_{1:T}}]$ and linearly projected via $\boldsymbol{\Phi}_{E}^{\text{proj}}$ to form a joint representation $E_{1:T}$:

\begin{equation}
\begin{aligned}
E_{t_{1:T}} &= \boldsymbol{\Phi}_{E}^{t}(A_{1:T}), \quad E_{f_{1:T}} = \boldsymbol{\Phi}_{E}^{f}(A_{1:T}), \\
E_{1:T} &= \boldsymbol{\Phi}_{E}^{\text{proj}}([E_{t_{1:T}} \parallel E_{f_{1:T}}]), \quad E_t \in \mathbb{R}^{256},
\end{aligned}
\end{equation}
where $E_{1:T}$ integrates both temporal dynamics and spectral characteristics for comprehensive emotion representation, and \([\cdot \parallel \cdot]\) represents the concatenation operation.

\subsection{Emotional Style Modeling Module}
\textbf{Emotional Styel Library Construction.}
Voiceprint information demonstrates a clear correlation with individuals, as it is shaped by unique physiological structures and neuromuscular control patterns, resulting in high individual specificity~\cite{reynolds2000speaker}. Based on this, for each subject, we use its voiceprint information as the base style representation. Specifically, there are \(K\) subjects in the training set, and each subject has \(N\) audio clips. We utilize a pre-trained speaker recognition model~\cite{bredin2020pyannote,Coria2020} to extract the voiceprint features of each audio clip. We average the voiceprint features of all audio clips for the same subject, obtaining the base style representation, denoted as \(\mathbf{R}_i\), where \(i = 1,2,\cdots,K\). Formally, we define:
\[
\mathbf{R}_i = \frac{1}{N} \sum_{j = 1}^{N} \mathbf{V}_{ij}, \quad \mathbf{R}_i \in \mathbb{R}^{512},
\]
Here, \(\mathbf{V}_{ij}\) represents the voiceprint feature of the \(j\)-th audio clip of the \(i\)-th subject.

However, the above base style representation is difficult to establish a connection with the emotional information of the audio. During the training process, each subject has \(C\) emotion categories, and each emotion category contains \(X\) audio clips. For each subject, we compute the average of the emotional features \(E_{1:T}\) across all audio clips belonging to the same emotion category. Subsequently, the averaged features is then concatenated with the base style representation \(\mathbf{R}_i\) to generate a personalized emotional style representation, which is denoted as \(\mathbf{P}_{ic}\), with \(c = 1,2,\cdots,C\). The formula is expressed as:
\[
\mathbf{P}_{ic} = [\mathbf{R}_i \parallel \frac{1}{X} \sum_{k = 1}^{X} \mathbf{E}_{ick}], \quad \mathbf{P}_{ic} \in \mathbb{R}^{768},
\]
Here, \(\mathbf{E}_{ick}\) represents the emotional feature of the \(k\)-th audio clip of the \(i\)-th subject under the \(c\)-th emotion category. In this way, we dynamically construct a two-dimensional \textit{\textless speaker $\times$ emotion\textgreater} emotional style library $\mathcal{L}$.

\textbf{Personalized Emotional Style Retrieval.}
Individuals with similar voiceprint characteristics exhibit convergent expressive styles, since voiceprint parameters encode both anatomical similarities and consistency in vocal habits~\cite{sabatier2019measurement}. During the retrieval process, for an input speech ${A}_{1: T}=\left({a}_{1}, \ldots, {a}_{T}\right)$, it is passed through $\boldsymbol{\Phi}_{E}$ and $\boldsymbol{\Phi}_S$ respectively to obtain the emotional feature \(E_{ic}\) and  voiceprint feature \(R_{i}\), which are then concatenated. The concatenated result is used to query in the pre-built emotional style library $\mathcal{L}$ to find the closest personalized emotional style representation. Specifically, the optimal style embedding is obtained via maximum similarity retrieval, which is expressed as:

\begin{gather}
S_{i} = \underset{S_k \in \mathcal{L}}{\arg\min}\, D_{\text{cos}}\big([E_{ic} \parallel R_{i}], S_k\big), \quad S_{i} \in \mathbb{R}^{768},\\
D_{\text{cos}}(u,v) = 1 - \frac{u^\top v}{\|u\|\|v\|}, \quad \text{where}\ u=[E_{ic} \parallel R_{i}],\ v=S_k,
\end{gather}
Here, $S_k$ represents an arbitrary emotional style feature in the emotional style library $\mathcal{L}$. The derived $S_{i}$ denotes the closest personalized emotional style to the input speech. Before being fed into subsequent processing stages, the retrieved features $S_{i}$ are passed through a linear projection layer to obtain a lower-dimensional representation.

\subsection{Emotion-content Disentanglement}
\textbf{Content Extractor.} We employ the pretained Wav2Vec 2.0 model~\cite{baevski2020wav2vec} as our content encoder $\boldsymbol{\Phi}_{C}$, building upon recent advances in speech-driven animation~\cite{fu2024mimic,emotalk,shen2024deitalk,fan2022faceformer}. The encoder processes raw audio ${A}_{1:T}$ through temporal convolutional layers $\boldsymbol{\Phi}_{C}^{conv}$ followed by a transformer encoder $\boldsymbol{\Phi}_{C}^{trans}$ with multi-head self-attention~\cite{vaswani2017attention}, and finally a linear projection layer $\boldsymbol{\Phi}_{C}^{proj}$ for dimensional adjustment. This pipeline generates frame-level content features:

\begin{equation}
C_{1:T} = \boldsymbol{\Phi}_{C} \left({a}_{1}, \ldots, {a}_{T}\right), \quad c_t \in \mathbb{R}^{256},
\end{equation}
We freeze $\boldsymbol{\Phi}_{C}^{conv}$ while fine-tuning both $\boldsymbol{\Phi}_{C}^{trans}$ and $\boldsymbol{\Phi}_{C}^{proj}$, maintaining stable acoustic feature extraction while adapting to our facial animation domain through learnable contextual modeling.

\begin{figure}[t]
\centering
\includegraphics[width=0.9\linewidth]{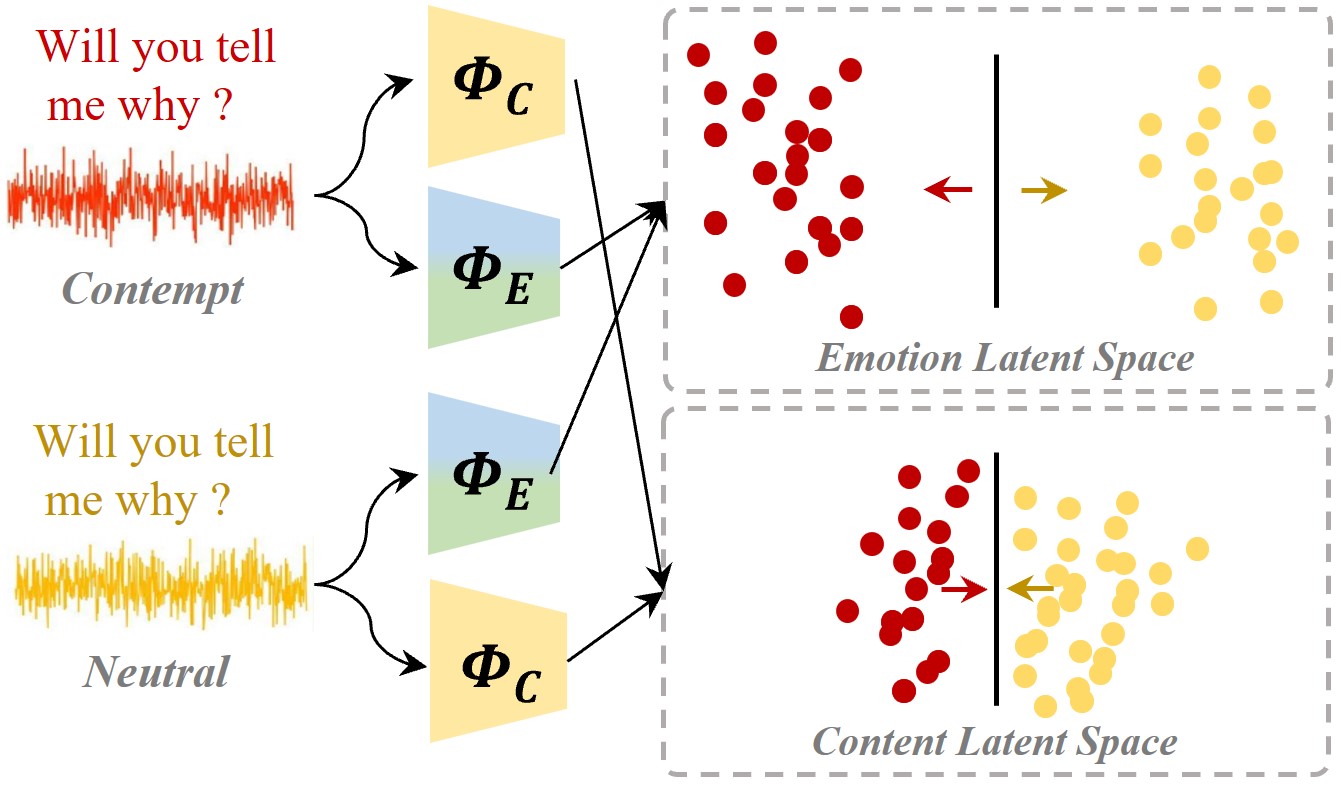}
\caption{Pairwise Disentanglement Mechanism. The content extractor $\boldsymbol{\Phi}_{C}$ and emotion extractor $\boldsymbol{\Phi}_{E}$ project paired audios (same semantics, different emotions) into content and emotion latent spaces, respectively. The mechanism pulls close semantically identical features while pushing apart emotionally distinct ones.  }
\Description{..}
\label{mechanism}
\end{figure}

\textbf{Pairwise Disentanglement Mechanism.} 
To enhance the model's ability to disentangle content features and emotional features, we proposes a new disentanglement mechanism (illustrated in Figure.~\ref{mechanism}). Compared with the strategy of forced cross-reconstruction adopted in previous work~\cite{emotalk,danvevcek2023emotional}, this mechanism does not require strict data pairing conditions and can achieve feature separation with a more concise architecture.  

Specifically, we input two audio clips that have the same content but differ in emotion, with one being neutral and the other emotional, denoted as $\{(A^{\text{neutral}}, A^{\text{emotion}})\}$. The content extractor and emotion extractor then project these paired audios into content and emotion latent spaces, respectively, pulling semantically identical features closer while pushing emotionally distinct ones apart. Formally:

\begin{equation}
\begin{aligned}
z_i^c &= \boldsymbol{\Phi}_C(A_i^{\text{neutral}}), \quad z_i^e = \boldsymbol{\Phi}_E(A_i^{\text{neutral}}), \\
z_j^c &= \boldsymbol{\Phi}_C(A_i^{\text{emotion}}), \quad z_j^e = \boldsymbol{\Phi}_E(A_i^{\text{emotion}}), \\
\mathcal{L}_{\text{dis}} &= \sum\nolimits_{i,j} \|z_i^c - z_j^c\|_{\psi} + \sum\nolimits_{i,j} \max(0, \delta - \|z_i^e - z_j^e\|),
\end{aligned}
\end{equation}
where $\psi$-norm enforces content compactness and $\delta$ denotes the learnable margin threshold. Through this gentle decoupling method, enabling the model to learn discriminative features that distinguish between emotional and neutral expressions. 

\subsection{Style-Guided Animation Generation}
The human face exhibits distinct motion patterns between upper and lower regions~\cite{kim2024deeptalk,shen2024deitalk}. The lower face displays high-frequency movements tightly coupled with speech articulation, while the upper face shows sustained, low-frequency motions reflecting emotional states. To model this biomechanical dichotomy, we propose a partitioned style-guided fusion decoder: a lower-face decoder $\boldsymbol{\Phi}_{D}^{low}$ that integrates phonetic content $C_{1:T}$, and personalized style $S_{1:T}$ for precise speech-synchronized animation; and an upper-face decoder $\boldsymbol{\Phi}_{D}^{up}$ that processes emotional $E_{1:T}$ and personalized style $S_{1:T}$ to generate emotion-driven facail motions. This specialized design enables accurate modeling of both speech-related articulations and sustained emotional expressions.

Both decoders employ an identical processing pipeline. First, periodic positional encoding ~\cite{fan2022faceformer} injects temporal regularity crucial for speech-synchronized lip movements. Next, a biased multi-head self-attention layer (inspired by ALiBi ~\cite{press2021train}) computes context-aware feature representations, where the attention bias mechanism prioritizes local temporal dependencies through its decaying mask pattern. The resultant features undergo transformation via feed-forward networks before final generation to facial blendshape. The architecture outputs 32 lower-face blendshapes and 20 upper-face blendshapes, which concatenate into 52-dimensional facial blendshape coefficients. 

\begin{table*}[h]
\centering
\caption{Quantitative evaluation results on blendshape-based datasets. Best performance highlighted in bold and lower values indicate better performance for metrics marked with $\downarrow$.}
\label{bs_results}
\begin{tabular}{lcccccccccccc}
\toprule
\multirow{2}{*}{Method} & 
\multicolumn{4}{c}{CREMA-D} & 
\multicolumn{4}{c}{RAVDESS} & 
\multicolumn{4}{c}{3D-EmoStyle} \\
\cmidrule(lr){2-5} \cmidrule(lr){6-9} \cmidrule(lr){10-13}
& {LBE$\downarrow$} & {PBE$\downarrow$} & {MBE$\downarrow$} & {BA$\uparrow$} & 
{LBE$\downarrow$} & {PBE$\downarrow$} & {MBE$\downarrow$} & {BA$\uparrow$} & 
{LBE$\downarrow$} & {PBE$\downarrow$} & {MBE$\downarrow$} & {BA$\uparrow$} \\
\midrule
EmoTalk~\cite{emotalk}                   & 0.208 & 0.287 & 0.374 & 0.713 & 0.166 & 0.242 & 0.371 & 0.752 & 0.199 & 0.259 & 0.313 & 0.642 \\
FaceDiffuser~\cite{stan2023facediffuser} & \textbf{0.168} & \textbf{0.254} & 0.753 & 0.742 & 0.152 & 0.266 & 0.416 & 0.721 & 0.178 & 0.218 & 0.458 & 0.627 \\
Ours                                     & 0.179 & 0.262 & \textbf{0.336} & \textbf{0.758} & \textbf{0.145} & \textbf{0.227} & \textbf{0.362} & \textbf{0.771} & \textbf{0.130} & \textbf{0.190} & \textbf{0.277} & \textbf{0.739} \\
\bottomrule
\end{tabular}
\end{table*}

\begin{table*}[h]
\centering
\caption{Quantitative evaluation results on vertex-based datasets.}
\label{mesh_results}
\addtolength{\tabcolsep}{-2pt}
\begin{tabular}{lccccccccc}
\toprule
\multirow{2}{*}{Method} & 
\multicolumn{3}{c}{CREMA-D} & 
\multicolumn{3}{c}{RAVDESS} & 
\multicolumn{3}{c}{3D-EmoStyle} \\
\cmidrule(lr){2-4} \cmidrule(lr){5-7} \cmidrule(lr){8-10}
& {LVE$\downarrow \scriptscriptstyle \times 10^{-3}$} & {EVE$\downarrow \scriptscriptstyle \times 10^{-4}$} & {FDD$\downarrow \scriptscriptstyle \times 10^{-4}$} & 
{LVE$\downarrow \scriptscriptstyle \times 10^{-3}$} & {EVE$\downarrow \scriptscriptstyle \times 10^{-4}$} & {FDD$\downarrow \scriptscriptstyle \times 10^{-4}$} & 
{LVE$\downarrow \scriptscriptstyle \times 10^{-3}$} & {EVE$\downarrow \scriptscriptstyle \times 10^{-4}$} & {FDD$\downarrow \scriptscriptstyle \times 10^{-4}$} \\
\midrule
EmoTalk~\cite{emotalk} & 4.829 & 8.605 & 4.633 & 6.823 & 8.081 & 4.469 & 3.544 & 9.848 & 4.565   \\
FaceDiffuser~\cite{stan2023facediffuser} & \textbf{3.871} & 9.562 & 4.806 & 3.423 & 9.351 & 4.865 & 3.498 & 14.486 & 8.196 \\
DEEPTalk~\cite{kim2024deeptalk} & 4.127  & 8.138 & 3.733 & 3.833 & 7.428 & 3.239 & 3.516 & 8.462 & 3.853  \\
Ours & 3.931 & \textbf{7.638} & \textbf{3.208} & \textbf{3.236} & \textbf{6.593} & \textbf{3.153} & \textbf{3.183} & \textbf{6.517} & \textbf{2.771} \\
\bottomrule
\end{tabular}
\end{table*}

\subsection{Objective function}
We utilize a comprehensive loss function that encompasses four key components: position loss, motion loss, classification loss, and disentanglement loss. The complete function is defined as follows:
\begin{equation}
\label{loss_all} 
L=\lambda_{1} L_{\text{pos}}+\lambda_{2} L_{\text{mot}}+\lambda_{3} L_{\text{cls}}+\lambda_{4} L_{\text{dis}},
\end{equation}
where $\lambda_{1}$ = 1, $\lambda_{2}$ = 0.5, $\lambda_{3}$ = 0.1, and $\lambda_{4}$ = 0.01 in all of our experiments. A comprehensive explanation of each of these components is provided in the following section. Please note that in the follow equations, $N$ denotes the number of samples in the training set, $T$ denotes the length of each input sequence.

\textbf{Position Loss.}
The position loss measures the difference between the predicted facial blendshape coefficients and the corresponding ground truth facial blendshape coefficients. Specifically, we use per-frame mean squared error (MSE) as the position loss:
\begin{equation}
\label{loss_pos} 
L_{\text{pos}}=\frac{1}{NT}\sum_{i}^{\boldsymbol{N}} \sum_{\mathbf{t}=1}^{\mathbf{T}}\left\|\left(
\hat{\boldsymbol{B}}_{1: T} - \boldsymbol{B}_{1: T}
\right)\right\|^{2}.
\end{equation}

\textbf{Motion Loss.}
To address jittery output frames, we introduce a motion loss that ensures temporal stability. This loss considers the smoothness between predicted and ground truth frames:
\begin{small}
\begin{equation}
\label{loss_mot} 
L_{mot}=\frac{1}{NT} \sum_{i}^{\boldsymbol{N}} \sum_{\mathbf{t}=1}^{\mathbf{T}}\left\|\left(\hat{\boldsymbol{B}_{2: T-1}} - \hat{\boldsymbol{B}_{1: T}}\right) - \left(\boldsymbol{B}_{2: T-1} - \boldsymbol{B}_{1: T}\right)\right\|^{2}.
\end{equation}
\end{small}

\textbf{Classification Loss.}
We introduce a classification loss to supervise the output of the emotion extractor $\boldsymbol{\Phi}_{E}$. The classification loss is defined as follows:

\begin{equation}
L_{cls}=-\frac{1}{N}\sum_{i}^{\boldsymbol{N}} \sum_{c=1}^{\boldsymbol{M}}\left(\boldsymbol{y}_{i c} * \log \boldsymbol{p}_{i c}\right),
\end{equation}
where $M$ denotes the number of classifications, $\boldsymbol{y}_{i c}$ is the observation function that determines whether the sample $i$ carries the emotion label $c$, and $\boldsymbol{p}_{i c}$ denotes the predicted probability that sample $i$ belongs to class $c$.

\textbf{Disentanglement Loss.} 
Our training framework processes content-parallel audio pairs $(A_{1:T}, \hat{A}_{1:T})$ with different emotional states—one neutral, one expressive. The model aims to maintain content via feature alignment and separate emotions via feature separation. The disentanglement mechanism is defined as:
\begin{equation}
\begin{aligned}
E &= \mathrm{Pool}(\boldsymbol{\Phi}_E(A_{1:T})), \quad \hat{E} = \mathrm{Pool}(\boldsymbol{\Phi}_E(\hat{A}_{1:T})), \\
C &= \mathrm{Pool}(\boldsymbol{\Phi}_C(A_{1:T})), \quad \hat{C} = \mathrm{Pool}(\boldsymbol{\Phi}_C(\hat{A}_{1:T})), \\
\mathcal{L}_{\text{dis}} &= 1 - \frac{1}{N}\sum_{i=1}^N \left[ \text{cos}(E_i, \hat{E}_i) - \text{cos}(C_i, \hat{C}_i) \right],
\end{aligned}
\end{equation}
where $\mathrm{Pool}(\cdot)$ aggregates frame-level features via average pooling, and $\text{cos}(\cdot,\cdot)$ measures cosine similarity.

\section{Experiments}
\textbf{Datasets.} 
\begin{figure}[h]
\centering
\includegraphics[width=0.9\linewidth]{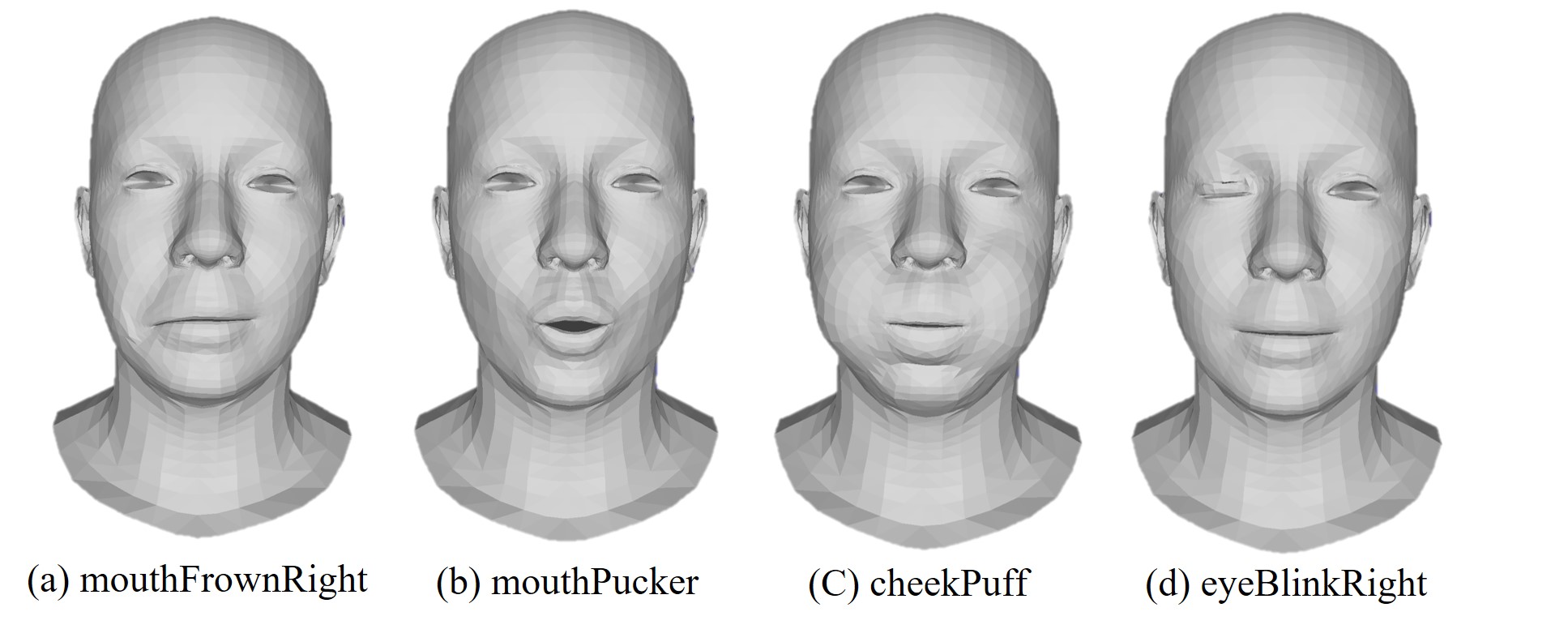}
\caption{Samples of reference meshes from facial blendshape coefficients to FLAME mesh.}
\label{flame}
\end{figure}
Due to the challenges in manually collecting high-quality emotional facial blendshape data from actors, we employ an established facial capture model~\cite{lugaresi2019mediapipe} to extract pseudo-blendshape coefficients from large-scale emotional 2D video datasets (MEAD~\cite{wang2020mead} and HDTF~\cite{zhang2021flow}), thereby constructing the 3D-EmoStyle dataset. This dataset comprises 180+ subjects exhibiting 8 prototypical emotional expressions. To facilitate comparative analysis with mesh-based approaches, we further convert the blendshape coefficients to mesh vertex data using the deformation transfer~\cite{sumner2004deformation}, generating mesh-based labels (shown in Figure~\ref{flame}). However, since MEAD's overlapping utterances between training and test sets compromise fair evaluation, we additionally incorporated CREMA-D~\cite{cao2014crema} and RAVDESS~\cite{livingstone2018ryerson} datasets for assessment. 

\textbf{Baseline Implementations.} 
Since our model aims to achieve personalized 3D facial emotion generation, we focus on comparing current emotion-based speech-driven methods for a fair comparison. Specifically, we conduct comprehensive comparisons between PESTalk and three state-of-the-art open-source approaches, EmoTalk~\cite{emotalk}, FaceDiffuser~\cite{stan2023facediffuser}, and DEEPTalk~\cite{kim2024deeptalk}. For DEEPTalk, we follow the methodology of the original paper by using EMOCAv2~\cite{danvevcek2022emoca} mdeol to extract FLAME parameters for model training. All models are trained on the training set of the 3D-EmoStyle dataset and subsequently evaluated on its test set along with two additional benchmark datasets.

\begin{figure}[h]
\centering
\includegraphics[width=\linewidth]{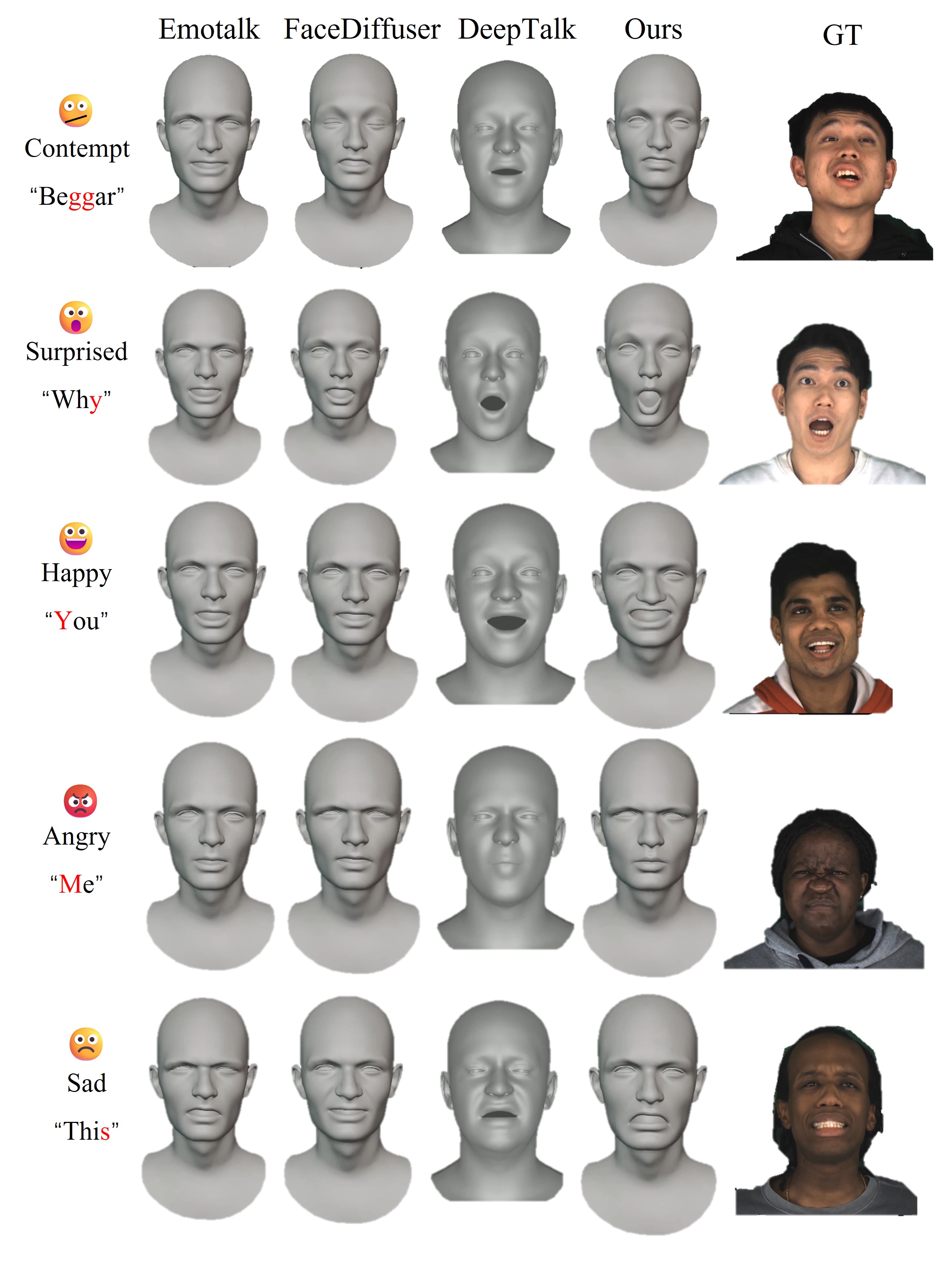}
\caption{Visual comparison of facial movements generated by different methods on the 3D-EmoStyle test set. The results demonstrate samples of distinct emotional expressions. Compared with other approaches, our method produces more emotionally expressive and realistic facial animations.}
\label{visual_mdeol}
\Description{..}
\end{figure}

\subsection{Quantitative Evaluation}
\textbf{Evaluation Metrics.} 
We employ a comprehensive set of evaluation using both blendshape-based and vertex-based metrics. For blendshape analysis, we adopt LBE (Lip Blendshape Error)~\cite{stan2023facediffuser} to measure lip-sync accuracy, while introducing PBE (Pronunciation Blendshape Error) to specifically evaluate pronunciation-related facial movements (\textit{e.g.} mouth and jaw). Additionally, MBE (Mean Blendshape Error)~\cite{stan2023facediffuser} is used to quantify the overall deviation across all blendshape parameters by computing the maximal L2 error per frame and averaging across the sequence. To assess temporal synchronization, we compute BA (Beat Alignment)~\cite{sun2024diffposetalk} between predicted and ground truth outputs. For vertex-based evaluation, we utilize three metrics: LVE (Lip Vertex Error)~\cite{fan2022faceformer,meshtalk} calculates the maximum L2 error among lip vertices to measure lip synchronization accuracy; EVE (Emotional Vertex Error)~\cite{emotalk} extends this evaluation to around the eyes and forehead regions for comprehensive expression assessment; and FDD (Upper Face Dynamics Deviation)~\cite{xing2023codetalker} compares the standard deviation of upper face vertex motions over time. 

\begin{figure}[h]
\centering
\includegraphics[width=0.9\linewidth]{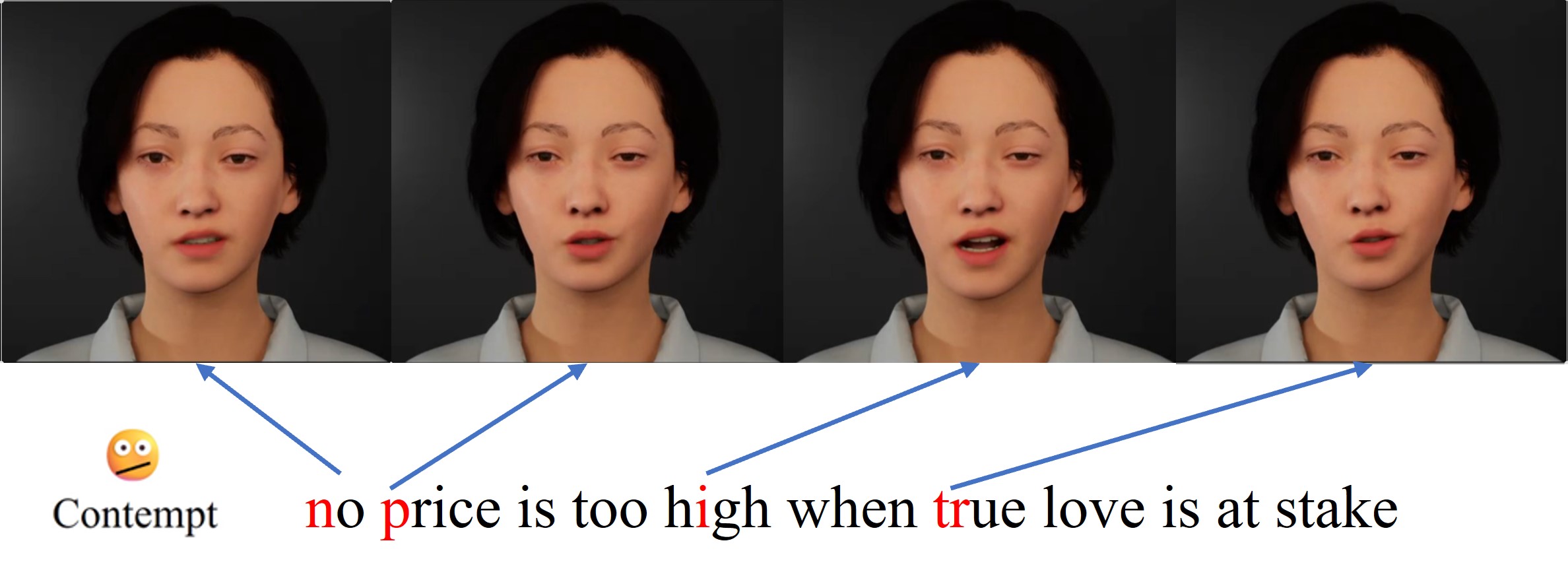}
\caption{Render results in Unreal Engine. The facial blendshapes generated by PESTalk are used to drive a digital human, which is then rendered in Unreal Engine.}
\label{bu1}
\Description{..}
\end{figure}

\textbf{Evaluation on Lip Synchronization.} As shown in Table~\ref{bs_results}, the PESTalk achieved remarkable performance across all tested datasets in three key metrics: LBE, PBE, and MBE. This robust performance demonstrates the model's strong capability of precise lip synchronization and generalization. Notably, PESTalk also attained the best scores in the BA metric. The BA metric evaluates synchronization by comparing audio beats with corresponding facial movement patterns derived from blendshape coefficients. Since lip motions occur at high frequencies, this can reflect the lip synchronization. These results further validate the exceptional lip-sync precision of PESTalk. Furthermore, while previous methods~\cite{meshtalk,fan2022faceformer,xing2023codetalker} often employ the LVE metric for lip-sync evaluation, our model directly outputs facial blendshape coefficients that cannot be directly compared. To address this, as presented in Table~\ref{mesh_results}, we converted the output blendshape coefficients into mesh vertex coordinates for comparison with ground truth data. Experimental results show that our model achieved the lowest LVE error across all datasets, providing additional compelling evidence for its accurate lip synchronization capability.

\textbf{Evaluation on Emotional Expression.} In evaluating emotional expression, we employed three key metrics: FDD to analyze discrepancies in upper facial movements compared to ground truth data, EVE to assess subtle expression variations in around the eyes and forehead regions (such as furrowed brows in anger expressions), and MBE to evaluate overall facial deviations. Experimental results demonstrate that PESTalk significantly outperforms all other methods across all datasets in terms of FDD, EVE, and MBE metrics (see Tables ~\ref{bs_results} and ~\ref{mesh_results}), conclusively validating its superior capability in emotional expression reproduction and its precision in capturing natural facial dynamics and nuanced expression details.

\begin{figure}[htbp]
\centering
\includegraphics[width=0.75\linewidth]{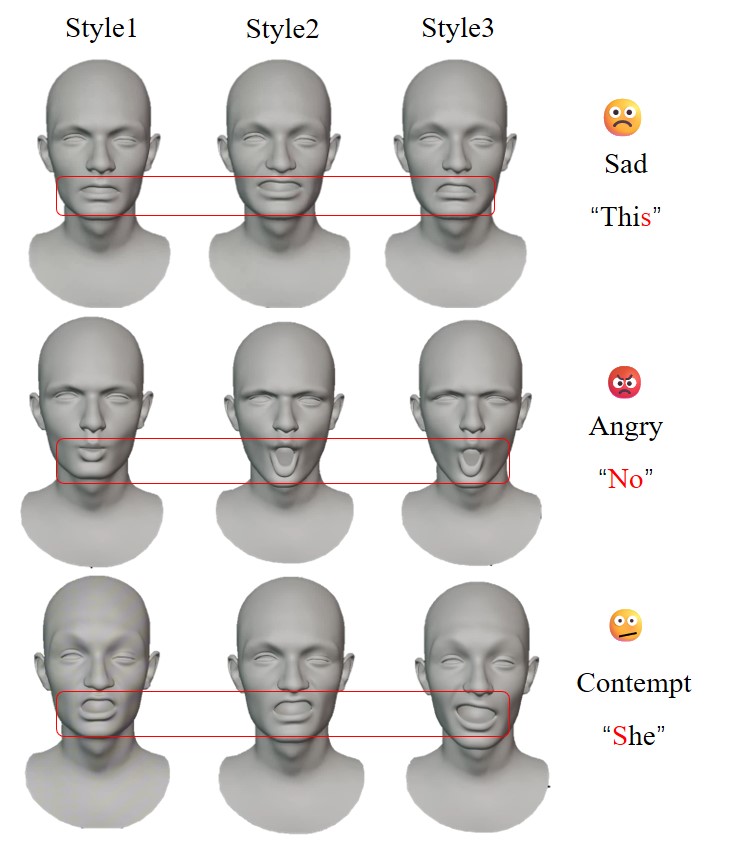}
\caption{The personalized emotional expression results of PESTalk. We evaluated the performance across samples of three distinct emotional states, where different individuals speaking the same sentence.}
\label{styles}
\Description{..}
\end{figure}

\subsection{Qualitative Evaluation}
\textbf{Visual Comparison.} 
In the comparative experiments shown in Figure~\ref{visual_mdeol}, we evaluated PESTalk against several state-of-the-art methods using proposed 3D-EmoStyle test set. The experimental results demonstrate that while most existing methods can generate relatively natural lip movements, they still exhibit significant shortcomings in detailed performance. Specifically, EmoTalk and FaceDiffuser show inadequate performance in speech synchronization accuracy. Taking the /$a\textsc{i}$/ phoneme in the word "why" as an example, these two methods fail to accurately reproduce the required articulatory features (the natural jaw drop and full mouth opening). In contrast, our model is capable of generating precisely lip-synced animations.

In terms of emotional expression, as shown in Figure~\ref{visual_mdeol}, it can be observed that except for DEEPTalk, other methods often produce incorrect or expressionless results that deviate from ground truth references. While DEEPTalk demonstrates improved emotional expressiveness, it still exhibits significant limitations. The method performs poorly in recognizing complex emotions such as "contempt," resulting in animations that substantially differ from the ground truth. In contrast, our approach consistently generates emotionally expressive and highly realistic facial animations across all emotional categories.

Additionally, PESTalk outputs facial blendshapes, enabling seamless integration into industry pipelines and enhancing usability for professional animators~\cite{emotalk,shen2024deitalk}. For instance, as illustrated in Figure~\ref{bu1}, we use MetaHuman4 to create our digital avatar and Unreal Engine 5 to render its animations. The animations are driven by the facial blendshape coefficients produced by our model. This approach allows our model to drive any digital character according to the rendering engine's rules, showcasing its strong practicality.

\textbf{Personalized Emotional Expression.} 
We constructed test groups comprising audio samples with identical semantic content and emotion category but from different speakers. As shown in Figure~\ref{styles}, the experimental results demonstrate that despite sharing identical emotion labels and textual content, PESTalk can accurately match personalized emotional expression styles based on speaker characteristics, generating highly realistic and individualized facial animations. This personalization is particularly in contempt expressions where noticeable differences existed in both the direction and degree of lip asymmetry among subjects.

\subsection{User Studies} 
To conduct a more comprehensive evaluation of our approach, we designed a user study with the following experimental protocol. Specifically, we randomly selected 5 samples from each emotion category of every subject in the 3D-EmoStyle test set. Then, we invited 20 participants to compare the results of all models with the ground truth. The participants evaluated the results in two dimensions (lip-sync accuracy and personalized emotional expressiveness) using a Mean Opinion Score (MOS) ranging from 1 to 5, where a higher score indicates better performance. As presented in Table~\ref {user_study}, the results clearly show that our method substantially outperforms existing approaches in terms of both lip-synchronization precision and personalized emotional expression.

\begin{table}[t]
\centering
\caption{User Study Results. Our method demonstrates superior performance compared to SOTA about personalized emotional expression (PEE) and lip synchronization (Lip Sync).}
\label{user_study}
\begin{tabular}{lcc}
\toprule
Method & PEE$\uparrow$ & Lip Sync$\uparrow$  \\
\midrule
EmoTalk~\cite{emotalk}                       & 3.878 & 3.598 \\
FaceDiffuser~\cite{stan2023facediffuser}     & 3.199 & 3.862  \\
DEEPTalk~\cite{kim2024deeptalk}              & 4.130 & 3.981 \\
Ours                                         & \textbf{4.349} & \textbf{4.276}  \\
\bottomrule
\end{tabular}
\end{table}

\begin{table}[t]
\centering
\caption{Ablation results on the 3D-EmoStyle test set.}
\label{ablation_study}
\begin{tabular}{lccccc}
\toprule
Method & LBE$\downarrow$ & MBE$\downarrow$ & {EVE$\downarrow \scriptscriptstyle \times 10^{-4}$ } & {FDD$\downarrow \scriptscriptstyle \times 10^{-4}$} \\
\midrule
w/o $\boldsymbol{\Phi}_{E}^{f}$    & 0.143  & 0.302 & 8.763 & 3.562 \\
w/o ESMM         & 0.149  & 0.317 & 9.202 &  4.103\\
w/o $\mathcal{L}_{\text{dis}}$     & 0.169  & 0.375 & 9.943 & 4.279 \\
Full (Ours)                        & \textbf{0.130} & \textbf{0.277} & \textbf{6.517} & \textbf{2.771} \\
\bottomrule
\end{tabular}
\end{table}

\subsection{Ablation Studies} 
We conducted comprehensive ablation studies on the 3D-EmoStyle test set to systematically evaluate the contributions of individual components in PESTalk. As demonstrated in Table~\ref{ablation_study}, removing the frequency stream of DESS (w/o $\boldsymbol{\Phi}_{E}^{f}$) significantly impairs the emotional expressiveness, evidenced by concurrent degradation in EVE and FDD. Similarly, the absence of ESMM compromises personalized emotional expression capabilities, which results in the output tending to be consistent for the same emotion category. Most notably, eliminating the disentanglement loss (w/o $\mathcal{L}_{\text{dis}}$) results in substantial performance drops across all metrics, as the model fails to effectively disentangle speech content from emotional features, thereby adversely affecting both lip-sync accuracy and emotional expression quality. 

\section{Conclusion} 
This paper introduces PESTalk, a novel speech-driven 3D facial animation framework that generates personalized emotional expressions while maintaining accurate lip synchronization. Unlike conventional approaches, PESTalk achieves this through two key components: (1) a Dual-Stream Emotion Extractor (DSEE) that captures both time-domain and frequency-domain emotional cues for fine-grained feature extraction, and (2) an Emotional Style Modeling Module (ESMM) that enables personalized emotional styles generation. To address data scarcity, we construct the 3D-EmoStyle, a large-scale 3D emotional talking face dataset with diverse personalized expressions. Extensive experiments on various datasets demonstrate the superiority of our model over existing methods across seven metrics, with notable improvements in emotional authenticity and personalized expression.

\begin{acks}
This work was supported by the Beijing Natural Science Foundation JQ23016, the Chinese National Natural Science Foundation Projects 62476273 and 62406320, the Science and Technology Development Fund of Macau Project 0123/2022/A3, 0044/2024/AGJ, 0140/2024/AGJ, and 0084/2024/RIB2.
\end{acks}

\bibliographystyle{ACM-Reference-Format}
\bibliography{sample-base}
\end{document}


\title{Supplementary Material}
\maketitle
In this supplementary material, we provide more details about PESTalk, which is comprised of three sections: 1) Detailed implementation aspects of PESTalk, including specifics on training parameters; 2) Comprehensive information on the creation of the 3D-EmoStyle dataset, covering data collection, preprocessing, and postprocessing; 3) Additional visualization experiments are conducted to verify the capabilities of PESTalk in emotion classification and personalized style generation.

\section{Implementation details}
PESTalk takes speech as input and generates a set of facial blendshape coefficients corresponding to the audio as output. The output parameters have a dimensionality of 52 and are sampled at a frame rate of 30 frames per second. For training, we use Adam Optimizer~\cite{kingma2014adam} with a batch size of 2 and a learning rate of 1e-4. The model is trained across four NVIDIA V100 GPUs, and the entire network requires approximately 40 hours to complete 150 epochs of training.

\section{Dataset construction details}
\textbf{Overview.} Our goal in constructing the dataset focuses on three core issues: 1) first, covering diverse emotional categories; 2) second, presenting expressions in multiple styles for a single emotion; 3) third, ensuring that the corpus of the dataset is rich.

\textbf{Acquisition scheme selection.} During the process of constructing the dataset, we have thoroughly considered various schemes for collecting facial blendshape coefficients. Currently, a relatively mature collection scheme on the market is to use the Live Link Face\footnote{https://apps.apple.com/us/app/live-link-face/id1495370836} to capture facial movements. Based on the ARKit\footnote{https://developer.apple.com/augmented-reality/arkit/} technology, this application can accurately track the movements of facial muscles and achieve real-time capture of facial expressions and movements. However, considering the requirement that our constructed dataset needs to cover multiple emotions and styles, the Live Link Face scheme sets extremely high standards for the acting ability of the actors, making it extremely difficult to implement in practice. It is worth noting that there are currently a large number of open-source 2D emotional video datasets~\cite{zhang2021flow,wang2020mead,cao2014crema,livingstone2018ryerson}. Based on this, we have explored a more efficient and concise scheme: using a facial motion capture model, Mediapipe~\cite{lugaresi2019mediapipe}, to extract 3D facial blendshape coefficients from a large number of 2D videos with emotions. We also compared the difference in acquisition accuracy between the two schemes. As shown in Figure~\ref{bu1}, by comparing the rendering results of the capture parameters of Live Link Face and Mediapipe, it can be found that the accuracies of the two are quite similar. 

\begin{figure}[h]
\centering
\includegraphics[width=0.7\linewidth]{figs/s_fig1.jpg}
\caption{Comparison of Live Link Face and Mediapipe rendering results. We randomly selected two frames corresponding to the animation results for comparison. The visual results indicate that both methods produce nearly identical outcomes, demonstrating comparable rendering performance.}
\label{bu1}
\Description{..}
\end{figure}

\textbf{Data Selection.}
The first two objectives of our dataset construction are essential for enabling the model to express emotions in a personalized manner, as the proposed Emotional Style Modeling Module (ESMM) relies heavily on the data within the database. Therefore, when constructing the dataset, it is necessary to cover different emotions and subjects, which should include people from various races and age groups. Based on this, we have selected the MEAD dataset~\cite{wang2020mead}. This dataset covers eight different emotions: angry, disgust, contempt, fear, happy, sad, surprise, and neutral. The actors are individuals aged between 20 and 35 from different races, and each person performs the same sentence with different emotions. However, MEAD has deficiencies in the richness of the corpus. Therefore, we have chosen the HDTF dataset~\cite{zhang2021flow} as a supplement. Its videos are sourced from YouTube and cover different speakers, scenes, and topics, providing a rich diversity in the corpus that reflects various real-world speaking contexts. The two datasets complement each other and jointly contribute to the construction of the high-quality dataset.

\textbf{Data processing.}
For all datasets, we select data captured from the front angle, as data from other angles do not facilitate accurate facial capture. Then, we standardize the frame rate of all videos to 30 frames per second and capture the facial blendshape for each frame. To further enhance the dataset's quality and minimize frame-to-frame jitter, we apply a Savitzky-Golay filter~\cite{schafer2011savitzky} with a window length of 5 and a polynomial order of 2 to the output facial blendshape coefficients. Once the process is complete, we render all the collected data and manually filter out any data with poor visual quality. This significantly improves the quality of the proposed dataset.

\begin{figure}[h]
\centering
\includegraphics[width=0.8\linewidth]{figs/s_fig2.jpg}
\caption{(a) Clustering results of emotion features for the 3D-EmoStyle test set. (b) Average expression style embeddings of ten randomly selected subjects across emotion categories in the emotional style library.}
\label{visual_embedding}
\Description{..}
\end{figure}

\textbf{Statistics.}
As a result, we develop the 3D-EmoStyle dataset, which comprises a total of \textbf{1,997,466} frames, \textbf{18.5} hours of data, \textbf{183} subjects, and \textbf{8,300} audio clips along with their corresponding facial 3D parameters. This dataset includes both male and female subjects, with ages ranging from \textbf{20 to 60} years. It features individuals of \textbf{different genders and ethnicities}, and encompasses \textbf{eight distinct emotional categories}.

\section{Visualization Experiments}
We conduct additional visualization experiments to validate PESTalk's emotional classification and personalized generation capabilities. As shown in Figure~\ref{visual_embedding} (a), we visualize the emotion features extracted by the dual-stream emotion extractor (DSEE) using t-SNE. The results clearly demonstrate that DSEE not only discriminates well between different emotion categories but also captures their unique characteristics accurately. This confirms that DSEE effectively models emotions, reinforcing the effectiveness of our approach in emotional expression generation.

Next, as shown in Figure~\ref{visual_embedding} (b), we randomly select expression style data for ten subjects from the emotional style library. For each subject, we compute the mean embedding across all emotion categories and apply dimensionality reduction using t-SNE for visualization. The results reveal significant inter-subject variations in emotional expression styles, providing strong evidence that our Emotional Style Modeling Module (ESMM) effectively captures individual differences, enabling the generation of personalized emotional expressions.

\bibliographystyle{ACM-Reference-Format}
\bibliography{sample-base}